# A switchable diode based on room-temperature two-dimensional ferroelectric α-In$_2$Se$_3$ thin layers

Siyuan Wan[1], Yue Li[1], Wei Li[2], Xiaoyu Mao[1], Wenguang Zhu[1,2,*] & Hualing Zeng[1,2,*]

1. International Center for Quantum Design of Functional Materials (ICQD), Hefei National Laboratory for Physical Science at the Microscale, and Synergetic Innovation Center of Quantum Information and Quantum Physics, University of Science and Technology of China, Hefei, Anhui 230026, China

2. Key Laboratory of Strongly-Coupled Quantum Matter Physics, Chinese Academy of Sciences, Department of Physics, University of Science and Technology of China, Hefei, Anhui 230026, China

* Corresponding author

**Abstract：Nanoscaled room-temperature ferroelectricity is ideal for developing advanced non-volatile high-density memories. However, reaching the thin film limit in conventional ferroelectrics is a long-standing challenge due to the possible critical thickness effect. Van der Waals materials, thanks to their stable layered structure, saturate interfacial chemistry and weak interlayer couplings, are promising for exploring ultra-thin two-dimensional (2D) ferroelectrics and device applications. Here, we demonstrate a switchable room-temperature ferroelectric diode built upon a 2D ferroelectric α-In$_2$Se$_3$ layer as thin as 5 nm in the form of graphene/α-In$_2$Se$_3$ heterojunction. The intrinsic out-of-plane ferroelectricity of the α-In$_2$Se$_3$ thin layers is evidenced by the observation of reversible spontaneous electric polarization with a relative low coercive electric field of ~2 × 10$^5$ V/cm and a typical ferroelectric domain size of around tens μm$^2$. Owing to the out-of-plane ferroelectricity of the α-In$_2$Se$_3$ layer, the Schottky barrier at the graphene/α-In$_2$Se$_3$ interface can be effectively tuned by switching the electric polarization with an applied voltage, leading to a pronounced switchable double diode effect with an on/off ratio of ~10$^4$. Our results offer a new way for developing novel nanoelectronic devices based on 2D ferroelectrics.**

**Introduction**

Ferroelectrics, featured with reversible spontaneous electric polarization, play an important role in modern electronics with novel functionality, such as high energy density capacitors (*1*), resistive switches (*2-5*), and non-volatile ferroelectric random-access memories (FERAMs) (*6-8*). Following the miniaturization of electronic devices, enormous efforts (*9-14*) are devoted on exploring ultra-thin or nanoscale ferroelectrics with the goal of realizing 2D ferroelectricity. However, the polarization instability in low-dimensional structure, for example the perovskite thin films (*10, 12, 15*), results in strong limit on their application in nanoelectronics. The origin of electric polarization vanishing below a critical size is complicated and still under debate (*16*). A generally accepted explanation is the presence of depolarization field in ultrathin ferroelectrics, which is due to the imperfect charge screening, local chemical environment, defects, and misfit strain at the interface (*10, 12, 13, 15, 17*). Therefore, overcoming these issues and realizing room-temperature stable ferroelectric phase in thin film limit are crucial for future development of ferroelectrics.

With innate stable layered structure and reduced surface energy, van der Waals (vdW) materials provide a versatile platform for studying 2D ferroelectricity at single atomic layer limit. Previous studies (*18-20*) have confirmed the possibility of seeking stable ferroelectric phase in layered material with inversion symmetry breaking. For example, Chang et al. (*19*) reported the in-plane ferroelectricity in lattice distorted monolayer SnTe at cryogenic temperature and Liu et al. (*20*) observed the out-of-plane electric polarization in 4 nm $CuInP_2S_6$ under ambient conditions. These pioneering findings inspire further study on layered 2D ferroelectrics. As a consensus, practical application of ferroelectricity relies on three aspects: the technically more important out-of-plane electric polarization relative to the film plane, high ferroelectric transition temperature ($T_c$) and simple lattice structure for material realization. Based on these above, a recent theoretical prediction (*21*) of intrinsic room-temperature 2D ferroelectricity with both in-plane and out-of-plane electric polarization in a widely studied vdW material,

α-In$_2$Se$_3$ (*22-28*), provides an ideal platform to explore the possibility of building novel 2D functional nanoelectronic devices.

Here, we report the observation of intrinsic out-of-plane ferroelectricity in α-In$_2$Se$_3$ thin layers and the demonstration of their practical ferroelectric device application under ambient conditions. Raman spectrum confirmed the crystal structure of ferroelectric In$_2$Se$_3$ thin layers to be alpha phase, which is consistent with previous theoretical prediction (*21*). Using piezoresponse force microscopy (PFM), we reveal the existence of room-temperature spontaneous ferroelectric polarization in 5 nm α-In$_2$Se$_3$. The two antiparallel electric dipoles in the out-of-plane direction were clearly characterized in the PFM phase image with a contrast of 180°. The single-point poling measurements on samples of different thicknesses suggest a switchable out-of-plane electric polarization and a low coercive electric field at the order of $10^5$ V/cm in α-In$_2$Se$_3$ thin layers. By applying an external electric field, we realize domain engineering with successful writing/erasing the polarity in a box-in-box pattern. Finally, as a demonstration of the device potential, we design and fabricate a ferroelectric diode in graphene/α-In$_2$Se$_3$/metal sandwich structure. The ferroelectric diode showed an on/off ratio of ~$10^4$. The current-voltage (I-V) hysteresis and a switchable diode effect have been observed. These findings in α-In$_2$Se$_3$ thin layers pave the way for developing novel 2D ferroelectric devices, such as non-volatile memory, high density capacitors, microelectromechanical systems (MEMS), and low power dissipation nanoelectronics.

**Results and Discussions**

**Crystal structure of α-In$_2$Se$_3$ thin layers.** In$_2$Se$_3$ is a semiconducting vdW material with direct band gap in the near infrared range (*18, 26*). It has been widely studied in the application of photo detectors (*18*), phase-change memory (*24*), and thermoelectric device (*28*). Despite intensive efforts, there are controversial identifications of its precise lattice structure between α and β phases (*23-26*). Therefore, characterizing and analyzing the crystal structure of α-In$_2$Se$_3$ is necessary before studying its ferroelectric property.

Figure 1 presents the schematic crystal structure of layered α-$In_2Se_3$. In a unit cell, there are three quintuple layers (QLs), which follow the ABC stacking (3R) sequence via weak vdW interactions. From the point of view of lattice symmetry, α-$In_2Se_3$ is explicitly asymmetry in structure (as can be seen in Fig. 1a), fulfilling the prerequisite of inversion symmetry breaking in ferroelectrics. Each α-$In_2Se_3$ QL contains five single-element (In or Se) monolayers. These monolayers are stacked in Se-In-Se-In-Se atomic layer sequence with strong covalence bonds. Within each In or Se atomic layer, the atoms form a triangular lattice. In an α-$In_2Se_3$ QL, the two outmost layers of Se atoms are both on the hollow sites of their neighboring layer of In atoms, with each Se atom being trigonal bonded to three In atoms and the same interlayer spacing. In contrast, the Se atoms in the middle layer are asymmetrically bonded to four In atoms from neighbor layers, constructing a tetrahedron structure with one vertical In-Se bond connected to one side while three In-Se bonds to the other side. As a consequence, there is dramatic difference in the interlayer spacings between the middle Se layer and the two In layers, which produces net electric dipoles in the out-of-plane direction (*21*). As shown in Fig. 1b, the out-of-plane ferroelectric polarization in α-$In_2Se_3$ has two degenerate states. The polarization direction, either upward or downward, is subjecting to the atomic position deviation of the middle-layer Se atoms from the inversion center of the QL and thereby can be switched by applying an external electric field.

With knowing its origin of ferroelectricity, we start with material realization. The α-$In_2Se_3$ thin layers were prepared by the well-established mechanical exfoliation method from bulk crystal (see Methods). Figure 2a shows typical optical image of α-$In_2Se_3$ nanoflakes on $SiO_2$/Si substrate with sharp optical contrast for different film thicknesses. Few-layer samples can be clearly visualized in the light/dark green area. The surface morphology and sample thickness were furthered characterized by Atomic force microscopy (AFM) in tapping mode under ambient condition. As shown in Figure 2b, the sample was atomically flat, indicating the characteristics of layered vdW materials. Some bright spots on the surface were residual polymers from the sticky tapes during exfoliation. Figure 2c shows the thickness profile measured along the red dash line in the sample topography image (Fig. 2b). The thinnest layer was found to be around 1.1 nm, which is closed to the previous reported thickness for α-$In_2Se_3$ monolayer (∼1 nm) (*25, 26, 29*). To the best of our knowledge, it is the first successful demonstration of achieving the high quality α-$In_2Se_3$ monolayer by mechanical exfoliation method from bulk.

To confirm the crystalline structure of α-$In_2Se_3$ thin layers studied in this work, we characterize our samples with Raman technique. Figure 2d shows a typical Raman spectrum taken from 50 nm α-$In_2Se_3$ thin layer with five dominant peaks. Similar Raman features were observed in thinner samples with worse signal to noise ratio. According to the phonon dispersion of ferroelectric α-$In_2Se_3$ by density function theory calculation (*21*), the unique features of Raman active modes for α phase $In_2Se_3$ are the softening of $A_1$ phonon mode (red shift if compared with β phase due to the structure phase transition) and the distinct optical phonon at ~260 cm$^{-1}$. In our Raman spectrum, we observed the E symmetry mode at 89 cm$^{-1}$ and the $A_1$(LO+TO) phonon mode at 104 cm$^{-1}$. By carefully checking the spectrum position of similar mode from β-$In_2Se_3$ in previous studies (*27, 30, 31*), centered at 110cm$^{-1}$, we found that the observed $A_1$(LO+TO) phonon showed slightly red shift. In addition, two peaks centered at 180 cm$^{-1}$ and 196 cm$^{-1}$ were attributed for $A_1$(LO) and $A_1$(TO) mode respectively, which were caused by the LO-TO splitting and confirmed the investigated $In_2Se_3$ to be asymmetry (belongs to R3m space group ) (*22*). In particular, we also observed the distinct Raman feature at 266 cm$^{-1}$, which could be regarded as the determinant evidence for the crystal structure of ferroelectric α-$In_2Se_3$. The observed Raman features shown in Fig. 2d exclusively confirm the α phase crystal structure of $In_2Se_3$ thin layers studied in this work.

**Ferroelectricity of α-$In_2Se_3$ thin layers.** Ferroelectrics are naturally with piezoelectric effect. Under an external electrical stimulus, the ferroelectric will deform either in phase (expand) or out of phase (contract) depending on the relative alignment between the polarization and the electric field（parallel or antiparallel）.Therefore, piezoresponse force microscope (PFM) is a versatile tool to study ferroelectricity. Figure 3 shows the typical PFM measurements from few layer α-$In_2Se_3$ with scanning area of 8×8 μ$m^2$ under ambient condition. The sample thickness was around 20 nm as shown in Fig. 3a. Spontaneous ferroelectric domains were directly visualized in the out-of-plane PFM phase image (Fig. 3b). By comparing the surface topography and the PFM phase image, we found there was no correlation between the ferroelectric domain structures and the morphology, which eliminated the possible origin of the phase contrast contributed from the variation of sample thickness. A 180º PFM phase contrast was observed, indicating the antiparallel ordering of electric dipoles in the out-of-plane direction between adjacent ferroelectric domains (Fig. 3c). The domain patterns were still discernible after several repeated measurements by PFM under room temperature. We also observed periodic ferroelectric stripe domains with a period of several nanometers in as-exfoliated α-$In_2Se_3$ (Supplementary Figure 1), which were commonly observed in $PbTiO_3$ (*32-34*). The antiparallel alignment of electric polarization between domains is the result of long-range electric dipole-dipole interactions with the lowest energy state as dictated by classical electrodynamics.

Another feature of ferroelectrics is the switch of electric polarization direction under an external electric field. To check this, we carried out local electric polarization switching measurements by varying the bias on the sample with the conductive PFM tip and the heavily doped Si substrate. Figure 3d and 3e show the out-of-plane amplitude (yellow) and PFM phase (green) as a function of the applied electric field. Clear single and butterfly-like electric hysteresis loops were observed in the phase and amplitude spectrum respectively. The 180º reversal of phase signal confirmed the characteristic switchable out-of-plane ferroelectric polarization in α-$In_2Se_3$ thin layers (*35*). The strongly nonlinear bias-dependent behavior in the PFM amplitude spectrum excluded the tip−sample electrostatic effect (*36*). In addition, the coercive field ($E_c$) of α-In$_2$Se$_3$ thin layer was found to be around 200 kV/cm (Supplementary Figure 2), which was far less than the reported value of 700 kV/cm for layered CuInP$_2$S$_6$ (*20*). The relative low $E_c$ in α-In$_2$Se$_3$ thin layer implies superior application in low-power 2D ferroelectric devices, such as the FERAMs. The measured PFM amplitude on ferroelectric domains allows the quantitative estimation of the longitudinal piezoelectric coefficient of α-In$_2$Se$_3$ films. The piezoelectric coefficient $d_{33}$, which represents the electromechanical response of α-In$_2$Se$_3$ thin layer in the c axis under an electric field in the same direction, can be calculated from

$$A=Qd_{33}V,$$

where *A* is the amplitude of piezo force response under the applied voltage *V* and *Q* (~24) is the quality factor of the resonance for the cantilever used in the measurement (Supplementary Fig. 3). The piezoelectric coefficient $d_{33}$ of 15 nm and 5 nm α-In$_2$Se$_3$ (Supplementary Fig. 3a & 3b) were estimated to be around $2\pm0.28$ pm/V and $1.17\pm0.38$ pm/V respectively. However, it should be noted that, due to the inhomogeneous distribution of the electric field at the PFM tip, there would be some error in the estimation of piezoelectric coefficient $d_{33}$ from PFM measurements. We further performed control experiments on BaTiO$_3$, hBN and doped Si substrate with the same technique (Supplementary Figure 4). Clear electric hysteresis loop was found in BaTiO$_3$ (BTO) single–domain strained films on SrTiO$_3$ substrate. The coercive field of 100 nm BTO was estimated to be 68 kV/cm, which was consistent with literature (*37*). The measurements from few-layer hBN and the heavily doped Si substrate showed paraelectricity without electric hysteresis loop, which ruled out the possible charge accumulation effect at the interfaces in the PFM measurements of α-In$_2$Se$_3$ thin layer.

To test the memory effect in ferroelectric α-In$_2$Se$_3$ thin layer, we artificially fabricated nanoscale ferroelectric domains with biased PFM tips. Figure 3f shows a 1.8×1.8 μm$^2$ out-of-plane PFM amplitude image with electrically written domain structure in α-In$_2$Se$_3$ thin layers of thickness down to 12 nm. The domain pattern was created by two steps. A 1×1 μm$^2$ area (denoted by a dashed-line frame in Fig. 3f) with either one of the two out-of-plane electric polarizations was firstly written under positive bias. Next, in the center, a small area of 0.5×0.5 μm$^2$ was rewritten or erased with a negative voltage. With these two steps, a box-in-box ferroelectric domain pattern was successfully created in the written/erased area. More importantly, there was no obvious damage on surface morphology in the corresponding area after writing/erasing the electric polarization. The retention performance of the ferroelectric domains in α-In$_2$Se$_3$ thin layer was found to be more than 24 hours, which can be found in Supplementary Figure 6.

**α-In$_2$Se$_3$ thin layers based ferroelectric diode.** As a demonstration of device potential, we design and build up a 2D ferroelectric diode with α-In$_2$Se$_3$/graphene vdW heterostructure. Figure 4a & 4b show the schematic device structure and the optical image of the device respectively. In the ferroelectric diode, the α-In$_2$Se$_3$ thin layers (> 5 nm as shown in Supplementary Fig. 7b) is sandwiched by p-type few layer graphene (FLG, thickness can be found in Supplementary Fig. 7b) and Aluminum as the bottom and top electrodes respectively (details can be found in Methods). Typical I-V curves of the ferroelectric diode show characteristic rectifying behavior as shown in Fig. 4c and Fig. 4d. The diodelike rectifying effect is due to the formation of metal-semiconductor junctions or Schottky barriers (details were discussed in Supplementary Note 3) within the two interfaces of our device (α-In$_2$Se$_3$/graphene and α-In$_2$Se$_3$/Al). However, unlike conventional diode, which conducts current in only one direction, the polarity of ferroelectric diode can be switched by the polarization state of α-In$_2$Se$_3$ thin layers.

By applying transient DC bias at $\pm 4$ V, which were high enough to polarize the ferroelectric α-In$_2$Se$_3$ thin layer, we found the rectifying behavior of the diode was switched. With +4 V bias triggering, the device showed forward rectifying behavior (Fig. 4c). In contrast, after applying -4V bias, the diode was reversed, showing backward rectifying effect (Fig. 4d). The unique switchable rectifying behavior of the ferroelectric diode is due to the modification of the Schottky barrier at the two interfaces by the following two aspects: (1) the generation of reversible built-in electric field in α-In$_2$Se$_3$ thin layer, which dramatically shifts the electronic bands with respect to vacuum level at its two sides (*4, 19*); and (2) the different doping in the bottom graphene electrode by the induced screening charges. The α-In$_2$Se$_3$ thin layers studied in this work is n-type. By considering the difference of its electron affinity and the working functions of top (Al) and bottom (p-type graphene) electrodes, we found that the Schottky barrier at α-In$_2$Se$_3$/graphene interface was much higher than that from α-In$_2$Se$_3$/Al interface (details can be found in Supplementary). As a result, the diode was originally conductive in the forward direction, from α-In$_2$Se$_3$ to the metal electrode. Fig. 4e shows the energy band diagram of our device at upward polarization state (+4 V triggering) with zero bias. The electric polarization of α-In$_2$Se$_3$ thin layers was upward (from bottom to top), producing strong built-in electric field pointing from the Al electrode to graphene electrode (downward). Thereby, the Schottky barrier at the graphene/α-In$_2$Se$_3$ interface was enhanced, while slightly reduced in the top interface between α-In$_2$Se$_3$ and the Al electrode, resulting enhanced forward rectifying behavior of the ferroelectric diode. In the case of -4 V polarized state, as seen from the band diagram in Fig. 4f, the built-in electric field was reversed. The graphene in the bottom was tuned to be n-type by the induced negative screening charges. As a consequence, the Schottky barrier at graphene/α-In$_2$Se$_3$ interface was strongly suppressed if compared to the upward polarized state. The device functioned as a reverse (backward) diode.

We further studied the I-V characteristic of the ferroelectric diode under high DC bias. The I-V curves (Supplementary Figure 7) were measured by sweeping the bias voltage from -4V to 4V and then backward to -4V. The sweeping rate was kept in relative low level to ensure the complete reversal of ferroelectric polarization. The I-V curves show distinct hysteresis behavior, indicating resistive switching effect (*4, 38*). Under high DC bias at both positive and negative side, due to the modification of Schottky barrier, the device is at its "on" state. By changing the same I-V curves with semilogarithmic scales (Supplementary Fig. 7e), we found the on/off ratio of the ferroelectric diode was at the order of ~$10^4$. The demonstration of ferroelectric diode based on α-In$_2$Se$_3$ thin layers pave the way for developing novel 2D functional devices.

**Conclusions**

In this work, we have demonstrated room-temperature stable out-of-plane ferroelectricity in α-In$_2$Se$_3$ thin layers with film thickness down to 5 nm. An external electric field can reverse the electric polarization and thereby modify the domain structure at nanoscale. In the graphene/α-In$_2$Se$_3$ heterostructure, we find a switchable diode effect, showing controllable rectifying I-V behavior. The ferroelectric α-In$_2$Se$_3$ provides a new platform for developing novel two-dimensional functional devices as well as the possibility of integration with other vdW materials.

**Methods**

**Sample preparation and device fabrication.** Atomically thin α-In$_2$Se$_3$ and few layer graphene were prepared by mechanical exfoliation method from single crystal bulk samples onto silicon wafers with 300 nm thick SiO$_2$ on top. The samples were firstly visualized with interference color contrast through optical microscope. Typical optical images can be found in Fig. 2a. The thickness was further confirmed by AFM (Bruker Dimension Icon) with tapping mode. The vdW heterostructures were stacked via all dry transfer technique (*39*). The devices were fabricated by standard optical lithography, with the electrodes made by deposition of 10 nm titanium and 100 nm aluminum via e-beam evaporator.

**Raman spectroscopy and PFM measurements.** Raman spectra were taken using Horiba Raman system (Labram HR Evolution) with 532 nm laser excitation. The on-sample power of the excitation was 100 μW. Piezoresponse force microscopy measurements were performed on Bruker Dimension Icon with Ir/Pt conducting tips (Nanoworld) with 300 kHz excitation frequency under ambient condition. The conductive substrates used were heavily doped silicon. Keysight B2900 source meter was used to measure the I-V characteristics. During sweeping process, the dwell time of every test point was 1 second to ensure that the reversal of electric polarization.

**Acknowledgments**

This work was supported by the National Key Research and Development Program of China (Grant No.2017YFA0205004 and 2017YFA0204904), the National Natural Science Foundation of China (Grant No.11674295, 11674299, 11374273, and 11634011), the Fundamental Research Funds for the Central Universities (Grant No. WK2340000082) and the China Government Youth 1000-Plan Talent Program. This work was partially carried out at the USTC Center for Micro and Nanoscale Research and Fabrication.

**Author Contributions**

H.Z. and W.Z. conceived the idea and supervised the research. S.W., W.L. and Y.L. prepared the samples and devices. S.W. carried out the PFM measurements and I-V characterization of ferroelectric diode. S.W. and H.Z. analyzed the data, wrote the paper, and all authors commented on the manuscript.

**Author Information**

The authors declare no competing financial interests. Correspondence and requests for materials should be addressed to Wenguang Zhu (wgzhu@ustc.edu.cn) and Hualing Zeng (hlzeng@ustc.edu.cn).

**Figures**

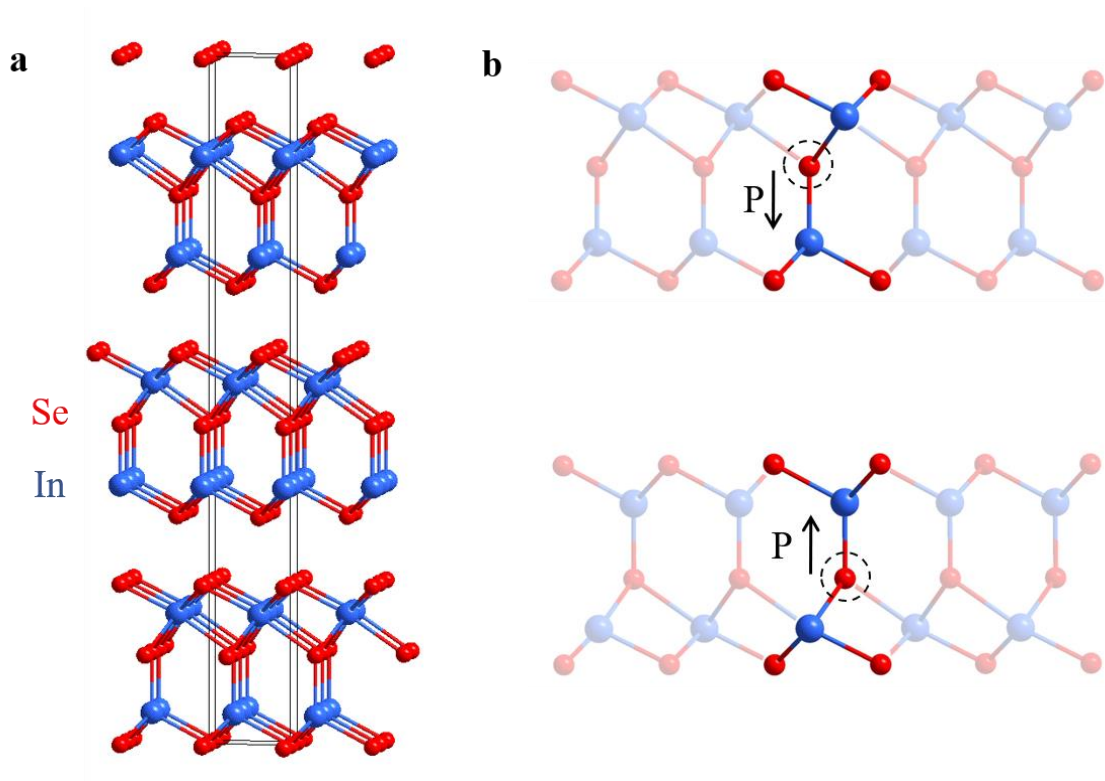

**Figure 1 Crystal Structure of α-In$_2$Se$_3$.** (**a**) Layer structure of α-In$_2$Se$_3$, with Indium atoms and Selenium atoms in blue and red respectively. Each quintuple layer (QL) contains five atomic layers in the order of Se-In-Se-In-Se layers. Three QLs form a unit cell. (**b**) Side views of the two oppositely polarized structures of one QL α-In$_2$Se$_3$. The direction of spontaneous out-of-plane electric polarization was indicated by black arrows.

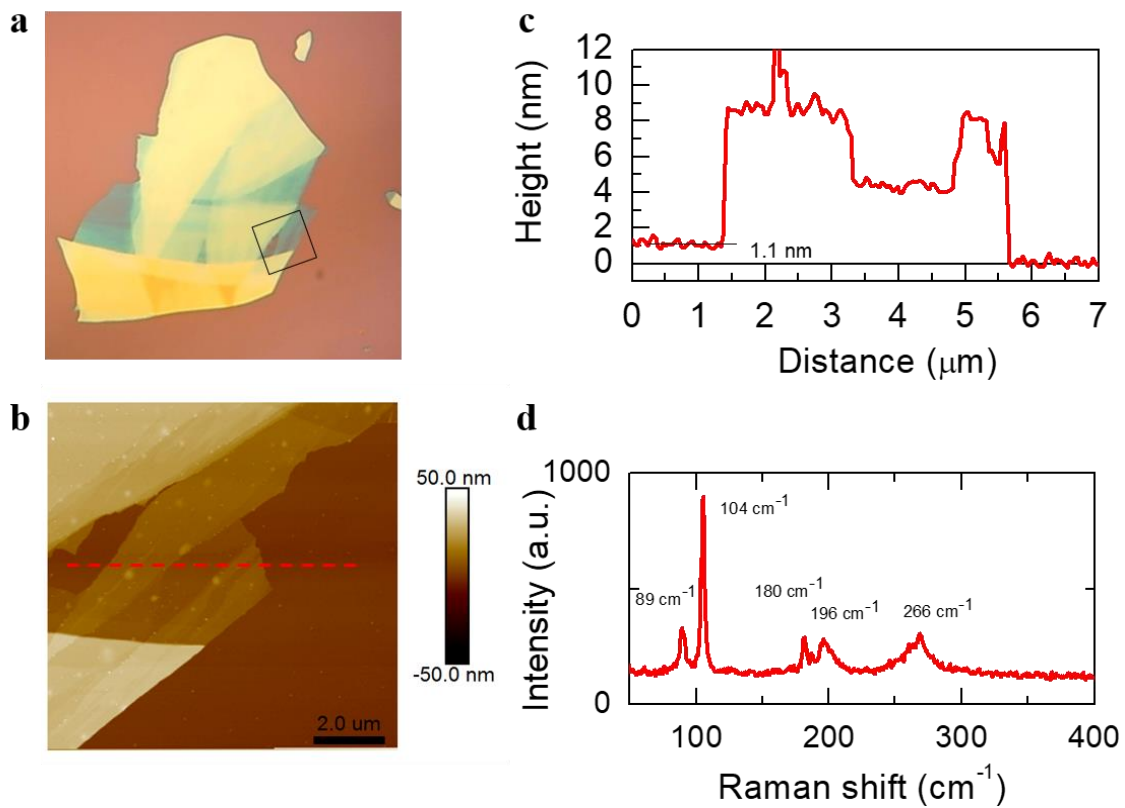

**Figure 2 Identification of α-In₂Se₃ thin layers.** (**a**) Optical image of mechanically exfoliated α-In$_2$Se$_3$ thin layers on SiO$_2$/Si substrate. (**b**) AFM topography (10×10 μm$^2$) of zoomed area as squared in **a**. (**c**) Corresponding height profiles taken along the red dash line in **b**. The thinnest film is 1.1 nm, closed to the thickness of monolayer ∼1 nm. (**d**) Raman spectrum of a 50 nm α-In$_2$Se$_3$ with 532 nm laser excitation. The four Raman peaks at 89, 104, 180 and 196 cm$^{−1}$ are consistent with previous reports. Distinct Raman feature for ferroelectric α-In$_2$Se$_3$ is identified to be at 266 cm$^{-1}$.

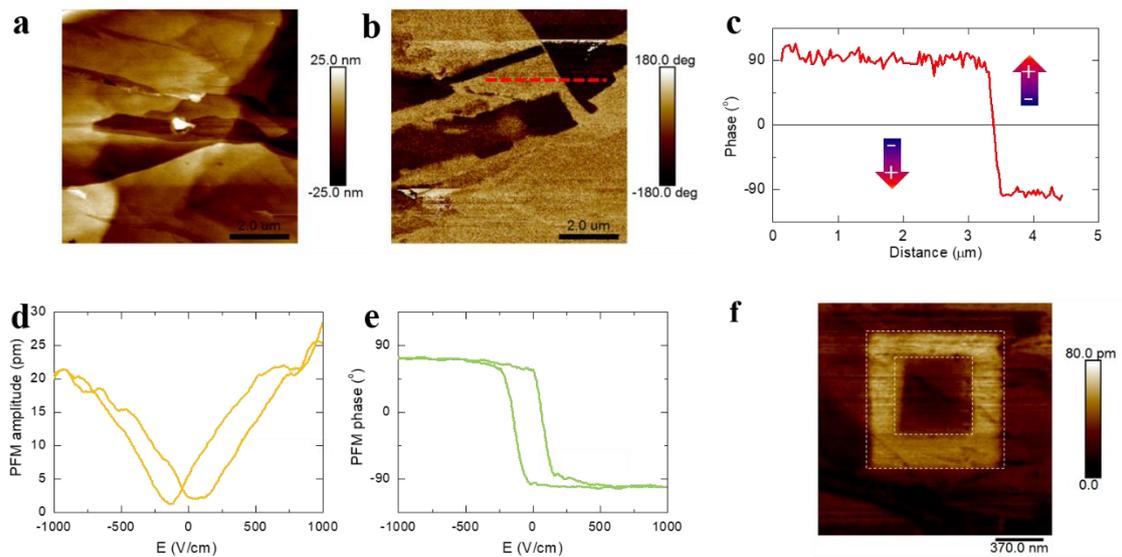

**Figure 3 Ferroelectricity of α-In$_2$Se$_3$ thin layers.** (**a**) The surface topography of α-In$_2$Se$_3$ thin layers (~20 nm) on heavily doped Si substrate. The scale bar is 2 μm. (**b**) The corresponding PFM phase image in out-of-plane direction, showing clear ferroelectric domains. (**c**) The phase profile of different ferroelectric domains as sketched by the red dash line in (**b**). A phase contrast of 180º is observed, which indicates the antiparallel directions of out-of-plane electric polarization between the adjacent domains. The arrows indicate the directions of electric polarization. (**d**) PFM amplitude and (**e**) PFM phase hysteresis loop measured from α-In$_2$Se$_3$ thin layers. (**f**) PFM amplitude image of domain engineering in α-In$_2$Se$_3$ with film thickness 12 nm. The scale bar is 370 nm.

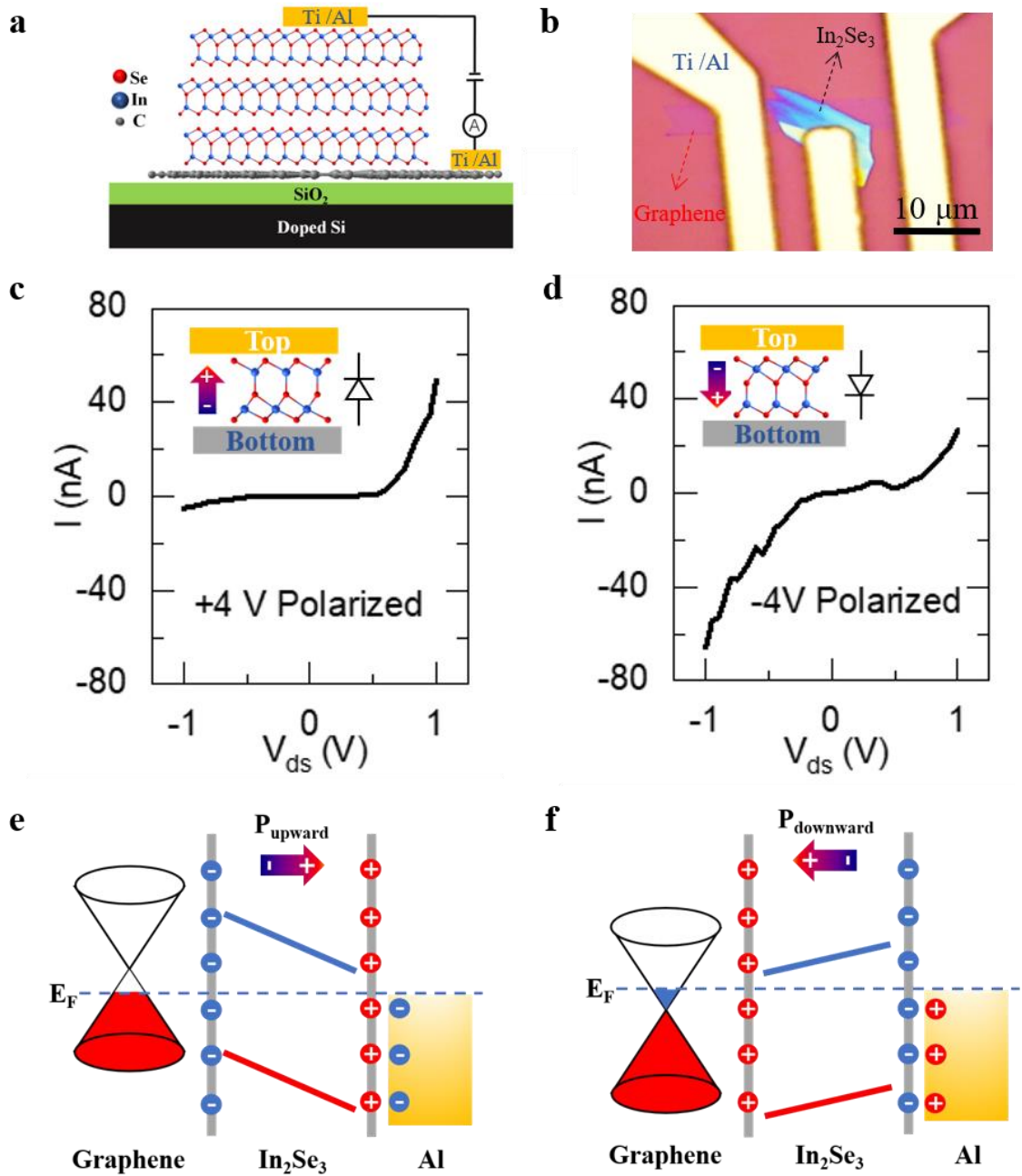

**Figure 4** Switchable ferroelectric diode based on α-In$_2$Se$_3$ thin layers. (**a**) Schematic and (**b**) optical image of the device. (**c**) and (**d**) I-V curves of ferroelectric diode with switchable rectifying behavior. (**e**) and (**f**) Schematic of energy band diagrams of graphene/In$_2$Se$_3$ heterostructure, illustrating the evolutions of Schottky barrier on the polarization state of the ferroelectric. The positive and negative charges on vertical grey lines stand for the polarization charges at the top and bottom sides of α-In$_2$Se$_3$ thin layer. The screening charges are visualized in the metal electrode.